\documentclass[12pt]{article}%
\usepackage{amssymb}
\usepackage{amsmath}
\usepackage{amsfonts}
\usepackage{graphicx}%
\providecommand{\U}[1]{\protect\rule{.1in}{.1in}}

\usepackage{hyperref}
\begin{document}
\bigskip\begin{titlepage}
		\vspace{.3cm} \vspace{1cm}
		\begin{center}
			\baselineskip=16pt \centerline{\bf{\Large{ Mimetic Dark Matter from Inflation}}}
			\vspace{1truecm}
			\centerline{\large\bf Ali H. Chamseddine$^{1,*}$\ , Viatcheslav Mukhanov$^{2,3,**}$ } \vspace{.5truecm}
			\emph{\centerline{$^{1}$Physics Department, American University of Beirut, Lebanon}}
			\emph{\centerline{$^{2}$Theoretical Physics, Ludwig Maxmillians University,Theresienstr. 37, 80333 Munich, Germany }}
			\emph{\centerline{$^{3}$School of Physics, Korea Institute for Advanced Study, Seoul 02455, Korea}}
		\end{center}
		\vspace{2cm}
		\begin{center}
			{\bf Abstract}
		\end{center}
	We investigate the coupling of mimetic dark matter to the Gauss-Bonnet
topological term in addition to the Einstein-Hilbert action. We demonstrate
that such interactions can naturally give rise to mimetic dark matter during
the inflationary stage of the universe's evolution. By choosing an
appropriate coupling between the mimetic field and the Gauss-Bonnet term, we
find that at the end of inflation, the correct amount of dust- like dark
matter is produced, with its energy density expressible in terms of the
Hubble parameter at the end of inflation. Furthermore, depending on the form
of the coupling, the post matter-radiation  equality behavior of mimetic
dark matter can experience slight modifications.\\[2em]
\href{mailto:}{*} {chams@aub.edu.lb}   {**}{Viatcheslav.Mukhanov@physik.uni-muenchen.de}
	\end{titlepage}

\section{Introduction}

A modest modification of gravity through the introduction of a constrained
field \cite{CM} has led to far-reaching consequences. In particular, the
modified equations admit solutions that effectively mimic an additional
dust-like matter component. Such solutions provide a compelling candidate for
explaining the observed dark matter in the universe, without invoking new
particles beyond the Standard Model.

Moreover, this field allows the construction of a diffeomorphism-invariant
combination involving the Laplacian of the mimetic field, which depends only
on the first derivatives of the metric and can be used to extend Einstein's
gravity in a simple and controlled manner \cite{mimcos}. In particular, the
mimetic field facilitates the resolution of long-standing singularity problems
in General Relativity, yielding non-singular cosmological \cite{Singular} and
black hole solutions \cite{BH}, \cite{AFmimetic}, \cite{BlackHoleRemnants}.
Importantly, these modifications preserve unitarity, as the graviton
propagator receives no contributions from higher time derivatives.

In our original work \cite{CM}, mimetic matter arose as an additional solution
characterized by an integration constant, whose fixed value could not change
over time. This immediately raised the question of what happens to this
mimetic cold dark matter during a sufficiently long inflationary stage. Taking
reasonable values for this constant, and assuming that dark matter does not
disrupt the exponential expansion, one finds that the amount of mimetic dark
matter remaining after inflation would be negligibly small.

To address this issue, we previously introduced a coupling between the mimetic
field and the inflaton; however, this approach was somewhat ad hoc and not
entirely convincing. In the present work, we propose a more natural mechanism.
Specifically, we explore the possibility of coupling the mimetic field to the
Gauss--Bonnet combination of curvature-squared terms, which, being
topological, does not affect the graviton propagator or violate unitarity. Our
aim is to determine whether such interactions can generate the observed amount
of dark matter during inflation purely in terms of the parameters of the
inflationary model. Furthermore, we investigate whether this coupling could
also influence the behavior of the dark energy component in the present epoch.

As a reminder, the mimetic field $\phi$ is introduced in order to isolate the
scale factor in $g_{\mu\nu}$ by constraining it to have a unit kinetic term
\begin{equation}
g^{\mu\nu}\partial_{\mu}\phi\,\partial_{\nu}\phi=1,\label{constraint}%
\end{equation}
where we assume signature $(+,-,-,-)$.

In the synchronous coordinate system
\begin{equation}
ds^{2}=dt^{2}-\gamma_{ij}(x^{k},t)\,dx^{i}dx^{j},\qquad
i,j=1,2,3,\label{eq:syn}%
\end{equation}
with $g_{00}=1$ and $g_{0i}=0,$ the solution of this constraint equation is
particularly simple:
\begin{equation}
\phi=t+A,\label{eq:sol}%
\end{equation}
where $A$ is a constant of integration. Hence, the mimetic field plays the
role of synchronous time. In these coordinates
\begin{equation}
-\nabla_{i}\nabla_{j}\phi=\kappa_{ij}=\frac{1}{2}\partial_{0}\gamma
_{ij}\label{eq:4}%
\end{equation}
which coincides with the extrinsic curvature of hypersurfaces of constant
time. The trace of the extrinsic curvature can be written in covariant form
as
\begin{equation}
\Box\phi=\kappa=\gamma^{ij}\kappa_{ij}=\frac{1}{\sqrt{\gamma}}\partial
_{0}\sqrt{\gamma},\label{eq:5}%
\end{equation}
where $\gamma=\det\gamma_{ij}$.

We now add to the Einstein--Hilbert action a Gauss--Bonnet term interacting
with $\phi$,%

\begin{equation}
\int d^{4}x\,\sqrt{-g}\,f(\phi)g\left( \Box\phi\right) \left( R_{\mu\nu
}^{\ \ \alpha\beta}R_{\alpha\beta}^{\ \ \mu\nu}-4R_{\mu\nu}R^{\mu\nu}%
+R^{2}\right) .
\end{equation}
In the synchronous gauge, the Gauss--Bonnet combination is topological and can
be written as a total derivative of a Chern--Simons three-form. As a result,
integration by parts implies that only the projection of this three-form onto
the three-dimensional spatial hypersurfaces contributes to the interaction.

The purpose of this paper is to study the influence of this term on the
expansion of the Universe. To make the presentation accessible to beginning
students, we include nontrivial step-by-step calculations in the Appendix.

\section{Gauss--Bonnet mimetic interactions}

We consider the simplest model of gravity with mimetic matter coupled to the
Gauss--Bonnet term:
\begin{equation}
S=\int d^{4}x\,\sqrt{-g}\left( -\frac{1}{2}R+\lambda\big(g^{\mu\nu}%
\partial_{\mu}\phi\,\partial_{\nu}\phi-1\big)+f(\phi)g\left( \Box\phi\right)
\left( R_{\mu\nu}^{\ \ \alpha\beta}R_{\alpha\beta}^{\ \ \mu\nu}-4R_{\mu\nu
}R^{\mu\nu}+R^{2}\right) \right) ,
\end{equation}
where we set $8\pi G=1.$ As shown in the Appendix, variation of this action
with respect to the metric yields the Einstein equation:
\begin{multline}
G_{\mu\nu}=2\lambda\,\partial_{\mu}\phi\,\partial_{\nu}\phi+\frac{2}%
{-g}\,R_{\kappa\lambda\gamma\delta}\,\nabla_{\beta}\nabla_{\alpha
}\!\big(f(\phi)g(\Box\phi)\big)\,\epsilon^{\sigma\beta\kappa\lambda}%
\epsilon^{\rho\alpha\gamma\delta}\,g_{\mu\rho}g_{\nu\sigma}\\
-\partial_{\nu}\phi\,\nabla_{\mu}\!\Big(f(\phi)g^{\prime}(\Box\phi
)\,GB\Big)-\partial_{\mu}\phi\,\nabla_{\nu}\!\Big(f(\phi)g^{\prime}(\Box
\phi)\,GB\Big)+g_{\mu\nu}g^{\alpha\beta}\nabla_{\alpha}\!\Big(\partial_{\beta
}\phi\,f(\phi)g^{\prime}(\Box\phi)\,GB\Big)\label{eq:7a}%
\end{multline}
where $GB$ denotes the Gauss--Bonnet term:
\begin{equation}
GB\equiv R_{\mu\nu}^{\ \ \alpha\beta}R_{\alpha\beta}^{\ \ \mu\nu}-4R_{\mu\nu
}R^{\mu\nu}+R^{2}\label{eq:GB-1}%
\end{equation}
and a prime denotes the derivative of the corresponding function with respect
to its argument.

Variation with respect to $\phi$ gives%

\begin{equation}
\frac{1}{\sqrt{-g}}\partial_{\mu}\Big(2\sqrt{-g}\,g^{\mu\nu}\partial_{\nu}%
\phi\,\lambda\Big)=f^{\prime}\left( \phi\right) g\left( \Box\phi\right)
GB+\frac{1}{\sqrt{-g}}\partial_{\mu}\Big(\sqrt{-g}\,g^{\mu\nu}\partial_{\nu
}\left( f\left( \phi\right) g\left( \Box\phi\right) GB\right)
\Big).\label{phi}%
\end{equation}
Finally, variation with respect to $\lambda$ yields the constraint
(\ref{constraint}).

Restricting to the flat Friedmann metric
\begin{equation}
ds^{2}=dt^{2}-a^{2}(t)\delta_{ij}\,dx^{i}dx^{j},
\end{equation}
we first note that, as follows from (\ref{eq:5}),
\begin{equation}
\Box\phi=\kappa=\frac{1}{\sqrt{\gamma}}\partial_{0}\sqrt{\gamma}=3\frac
{\dot{a}}{a}.\label{eq:11}%
\end{equation}
Calculating the components of Riemann tensor:
\[
R_{0i0j}=\ddot{a}a\,\delta_{ij},\quad R_{ijkl}=-\dot{a}^{2}\left(  \delta
_{ik}\delta_{jl}-\delta_{il}\delta_{jk}\right)
\]
and its contractions:
\begin{align}
R_{00} &  =R_{0i0}^{\ \ \ i}=-3\frac{\ddot{a}}{a},\\
R_{ij} &  =R_{i0j}^{\ \ \ 0}+R_{ikj}^{\ \ \ k}=\left(  \frac{\ddot{a}}%
{a}+2\left(  \frac{\dot{a}}{a}\right)  ^{2}\right)  a^{2}\delta_{ij},\\
R &  =R_{00}-\frac{1}{a^{2}}\delta^{ij}R_{ij}=-6\left(  \frac{\ddot{a}}%
{a}+\left(  \frac{\dot{a}}{a}\right)  ^{2}\right)  .
\end{align}
we find the Gauss-Bonnet term
\begin{equation}
GB=R_{\mu\nu\rho\sigma}R^{\mu\nu\rho\sigma}-4R_{\mu\nu}R^{\mu\nu}%
+R^{2}=24\,\frac{\ddot{a}\dot{a}^{2}}{a^{3}}.\label{eq:GB}%
\end{equation}
Notice that
\begin{align}
\sqrt{-g}GB &  =24\overset{..}{a}\left(  \overset{.}{a}^{2}\right)
=8\partial_{0}\left(  \overset{.}{a}^{3}\right) \nonumber\\
&  =\partial_{0}\left(  CS\right)
\end{align}
so that the Chern-Simons (CS) term is
\begin{equation}
CS=8\left(  \overset{.}{a}^{3}\right)
\end{equation}
For a flat Friedmann universe $G_{00}$ equation (\ref{eq:7a}) becomes
\begin{align}
3\left(  \frac{\dot{a}}{a}\right)  ^{2}  & =2\lambda-24\,\partial
_{0}\!\big(f(t)g(\kappa)\big)\,\frac{\dot{a}^{3}}{a^{3}}-48\,\partial
_{0}\!\left(  f(t)g^{\prime}(\kappa)\frac{\ddot{a}\,\dot{a}^{2}}{a^{3}}\right)
\nonumber\\
& +24\left(  \partial_{0}+3\frac{\dot{a}}{a}\right)  \left(  f(t)g^{\prime
}(\kappa)\frac{\ddot{a}\,\dot{a}^{2}}{a^{3}}\right)  .\label{eq:00eq}%
\end{align}
The equation (\ref{phi}) simplifies to:
\begin{equation}
\partial_{0}\Big(2a^{3}\,\lambda\Big)=24f^{\prime}\left(  t\right)  g\left(
\kappa\right)  \ddot{a}\dot{a}^{2}+\partial_{0}\left(  a^{3}\partial
_{0}\left(  f\left(  t\right)  g\left(  \kappa\right)  24\,\frac{\ddot{a}%
\dot{a}^{2}}{a^{3}}\right)  \right) \label{eq:lmeq}%
\end{equation}
which can be integrated to give
\[
2\lambda=\frac{8}{a^{3}}\int dt\,f^{\prime}(t)\,g(\kappa)\,\partial_{0}\left(
\dot{a}^{3}\right)  +24\,\partial_{0}\!\left(  f(t)g^{\prime}(\kappa
)\frac{\ddot{a}\,\dot{a}^{2}}{a^{3}}\right)
\]
Substituting this expression into (\ref{eq:00eq}) and noting that
\begin{equation}
\frac{\ddot{a}}{a}=\frac{1}{3}\dot{\kappa}+\frac{1}{9}\kappa^{2},\label{eq:ka}%
\end{equation}
where $\kappa=3H\equiv3\dot{a}/a$ (\ref{eq:11}), we can simplify the 0-0
Einstein equation:
\begin{equation}
\frac{1}{3}\kappa^{2}=-\frac{24}{a^{3}}\int\dot{a}\,\partial_{0}(\dot{a}%
^{2}f^{\prime}\left(  t\right)  g\left(  \kappa\right)  )\,dt+\frac{8}%
{27}\kappa^{5}f\left(  t\right)  g^{\prime}\left(  \kappa\right)
+\varepsilon,\label{eq:master}%
\end{equation}
where $\varepsilon$ represents the contribution of ordinary matter. The
indefinite integral here reflects the existence of mimetic dark matter.

\section{Inflation and Dark Matter}

Let us assume that at the end of inflation, the behavior of the metric is
determined by the slowly varying potential of the scalar field, $\varepsilon
\approx V\left( \varphi\right) $. The Hubble parameter is approximately
constant during inflation, $H_{I}\approx\sqrt{V/3}$, and the scale factor
evolves as
\begin{equation}
a\approx a_{f}\exp H\left( t-t_{f}\right) ,\label{eq:sf}%
\end{equation}
where $a_{f}$ is the scale factor at the end of inflation $\left(
t=t_{f}\simeq1/H_{I}\right) $.

In most inflationary scenarios, after inflation, the inflaton behaves like
massive non-relativistic particles, so that during this phase $a\propto
t^{2/3}$ for $t_{f}<t<t_{rad}$, until reheating occurs. After reheating, these
particles decay into ultra-relativistic particles, and the Universe enters the
radiation-dominated era at $t>t_{rad}$, when $a\propto t^{1/2}$. This
radiation-dominated stage continues until $t_{eq},$ when the energy density of
radiation becomes comparable to that of cold matter, after which cold matter
dominates again. Taking this into account, the energy density of the inflaton,
which at the end of inflation was $\varepsilon=3H_{I}^{2}$, after decaying
into radiation $\left(  t>t_{rad}\right)  $ scales as
\begin{equation}
\varepsilon=3H_{I}^{2}\frac{a_{f}^{3}a_{rad}}{a^{4}},\label{eq:rad}%
\end{equation}
Let us assume that the function $f$ is linear in $\phi$, i.e., $f\left(
\phi\right)  =\beta\phi$ and $g\left(  \kappa\right)  =\kappa^{3}=27H^{3}$. To
calculate the integral in equation (\ref{eq:master}) from the beginning of
inflation until $t>t_{rad}$, it is convenient to rewrite it as
\begin{equation}
\int_{t_{in}}^{t}\dot{a}\,\partial_{0}(\dot{a}^{2}f^{\prime}g)\,dt=27\beta
\int_{0}^{a}\frac{d}{da}\left(  a^{2}H^{5}\right)  \,aH\,da,\label{eq:int}%
\end{equation}
During inflation, where $H=H_{I}$ is approximately constant, the contribution is%

\begin{equation}
27\beta\int_{0}^{a_{f}}\frac{d}{da}\left(  a^{2}H^{5}\right)  \,aH\,da=18\beta
H_{I}^{6}a_{f}^{3}.
\end{equation}
After inflation, during the cold inflaton particle domination and subsequent
radiation domination, the Hubble parameter evolves as
\begin{equation}
H\left(  a\right)  =H_{I}\left(  \frac{a_{f}}{a}\right)  ^{3/2},\,\;\quad
H\left(  a\right)  =H_{I}\frac{a_{f}^{3/2}a_{rad}^{1/2}}{a^{2}},
\end{equation}
correspondingly. Accordingly, we find
\begin{equation}
27\beta\int_{a_{f}}^{a}\frac{d}{da}\left(  a^{2}H^{5}\right)  aHda=27\beta
H_{I}^{6}\left(  -\frac{11}{12}a_{f}^{3}+\frac{1}{36}\frac{a_{f}^{9}}%
{a_{rad}^{6}}+\frac{8}{9}\frac{a_{f}^{9}a_{rad}^{3}}{a^{9}}\right)  .
\end{equation}
where we have split the integral $\int_{a_{f}}^{a}\left(  \cdot\right)
=\int_{a_{f}}^{a_{rad}}\left(  \cdot\right)  +\int_{a_{rad}}^{a}\left(
\cdot\right)  .$ Combining these results, equation (\ref{eq:master}) for
$t>t_{rad} $ can be written as
\begin{equation}
3H^{2}=162\beta H_{I}^{6}\frac{a_{f}^{3}}{a^{3}}\left(  1-\frac{1}{9}\left(
\frac{a_{f}}{a_{rad}}\right)  ^{6}\right)  +396\beta H_{I}^{6}\frac{a_{f}%
^{9}a_{rad}^{3}}{a^{12}}+3H_{I}^{2}\frac{a_{f}^{3}a_{rad}}{a^{4}%
}.\label{eq:main}%
\end{equation}
It is clear that reheating requires some time after inflation. The transition
to the radiation era occurs at
\begin{equation}
t_{rad}\simeq Nt_{f}\simeq N/H_{I},
\end{equation}
with $N\sim20-100$ (see, e.g. ADD \cite{Book}). Noting that the second term in
the bracket on the right-hand side is negligible and that the term decaying as
$a^{-12}$, can be dropped, equation (\ref{eq:main}) simplifies to
\begin{equation}
3H^{2}=162\beta H_{I}^{6}\frac{a_{f}^{3}}{a^{3}}+3H_{I}^{2}\frac{a_{f}%
^{3}a_{rad}}{a^{4}}.
\end{equation}
Here, the first term represents the mimetic dark matter, while the second term
is the contribution of radiation from the decay of the inflaton. These two
contributions become equal at the time of equality, $t_{eq}$, when $a=a_{eq}%
$:
\begin{equation}
\frac{a_{eq}}{a_{rad}}=\frac{1}{54\beta}H_{I}^{-4}.\label{eq:ratio}%
\end{equation}
Taking into account that during the radiation-dominated stage $a\propto
t^{1/2}$, we obtain
\begin{equation}
t_{eq}\approx\left(  \frac{N^{1/2}}{54\beta}\right)  ^{2}H_{I}^{-9}%
\end{equation}
in Planck units, where $8\pi G=1$ and $t_{Pl}=2.7\cdot10^{-43}\sec.$

As an example, taking $H_{I}\sim10^{-6}$ in Planck units, $N\sim100$ and
$\beta\sim0.1,$ we find
\[
t_{eq}\sim10^{12}\sec
\]
in agreement with observations.

As it is clear from the consideration above that the energy density of mimetic
matter during inflation remains nearly constant and determined by the Hubble
constant. Therefore after decay of the inflaton field the generated
perturbations are adiabatic.

\section{Anomalous Dark Matter}

By considering nonlinear functions $f\left(  \phi\right)  $, one can obtain
anomalous behavior of mimetic cold dark matter during the stage of dark matter
domination. While many models can exhibit similar behavior, for simplicity and
to illustrate the idea, we consider the simplest case:
\begin{equation}
f\left(  \phi\right)  =-\frac{\alpha}{16}\phi^{2},\quad g=1.
\end{equation}
At the stage of mimetic matter domination, we neglect the contribution of
other matter components. Using $\phi=t$ and setting $\varepsilon=0$, the
master equation (\ref{eq:master}) becomes:
\begin{equation}
H^{2}=\frac{\alpha}{a^{3}}\int^{t}\dot{a}\,\partial_{0}(\dot{a}^{2}%
t)dt.\label{eq:simplified}%
\end{equation}
To solve this equation, we adopt the ansatz $a\propto t^{n}$ and assume
$n>2/3$. In this case, the main contribution to the integral comes from the
upper limit, and the equation (\ref{eq:simplified}) reduces to a quadratic
equation for $n$:
\begin{equation}
n^{2}-\frac{1}{2}\left(  1+\frac{3}{\alpha}\right)  n+\frac{1}{\alpha
}=0,\label{eq:quad}%
\end{equation}
whose solutions are
\begin{equation}
n=\frac{1}{4}\left(  1+\frac{3}{\alpha}\right)  \left(  1\pm\sqrt
{1-\frac{16\alpha}{\left(  3+\alpha\right)  ^{2}}}\right)  .\label{eq:36}%
\end{equation}
The expression under square root is positive only for $\alpha<1$ and
$\alpha>9.$ Expanding the solution with the negative sign in front of the
square root to first order in $\alpha$ we obtain:
\begin{equation}
n\simeq\frac{2}{3}\left(  1+\frac{\alpha}{9}\right)  .
\end{equation}
Consequently, the relation between the Hubble parameter and cosmic time is
modified as:
\begin{equation}
H=\frac{2}{3}\left(  1+\frac{\alpha}{9}\right)  \frac{1}{t}.
\end{equation}
The analysis of other interesting cases of anomalous behavior of mimetic
matter is left for the reader.

\section{Conclusions}

It is now well established that the constraint (\ref{constraint}) on a field
$\phi$, combined with a longitudinal part of the metric, mimics cold dark
matter. In synchronous coordinates, $\phi$ represents time. This modest
modification of General Relativity provides a simple explanation for dark
matter without introducing new particles, and allows higher-derivative
interactions without generating ghost or tachyonic modes in the graviton propagator.

Moreover, since $\Box\phi=\kappa$ corresponds to the trace of the extrinsic
curvature of synchronous constant-time hypersurfaces, one can use the first
time derivative of the metric---introduced in a covariant way---to modify the
Einstein action via terms $f(\Box\phi)$. By a suitable choice of this
function, one can resolve singularities in Friedmann and Kasner universes, as
well as inside black holes \cite{BlackHoleRemnants,AFmimetic}. Furthermore,
the mimetic field allows the Horava gravity to be reformulated in a fully
covariant manner \cite{MimeticHorava}. The mimetic modification can also be
used to avoid the self-reproduction problem in cosmology \cite{CKV}.

In this paper, we have investigated the consequences of coupling the mimetic
field with the topological Gauss--Bonnet curvature invariant. We have shown
that, in this scenario, mimetic cold dark matter can be naturally generated
during the inflationary stage, with a density determined entirely by the
Hubble parameter during inflation, and consistent with current observations.
Moreover, we have demonstrated that coupling the mimetic field with the
Gauss--Bonnet term can induce an interesting anomalous behavior of mimetic matter.

\section*{Acknowledgements}
\addcontentsline{toc}{section}{Acknowledgements}

The work of A.~H.~C.\ is supported in part by the National Science Foundation
Grant No.\ Phys-2207663. He would also like to acknowledge hospitality of the
Korea Institute for Advanced Study, Seoul, where this work was partially done.

\section{Appendix: Equations of motion}

To vary the action, note that the Gauss--Bonnet term is topological and can be
written as
\begin{align}
\int\epsilon_{abcd}R^{ab}\wedge R^{cd} &  =\frac{1}{4}\int\epsilon
_{abcd}R_{\mu\nu}^{\ \ ab}R_{\kappa\lambda}^{\ \ cd}\,dx^{\mu}\wedge dx^{\nu
}\wedge dx^{\kappa}\wedge dx^{\lambda}\nonumber\\
&  =\frac{1}{4}\int d^{4}x\,\epsilon_{abcd}R_{\mu\nu}^{\ \ ab}R_{\kappa
\lambda}^{\ \ cd}\,\epsilon^{\mu\nu\kappa\lambda}\nonumber\\
&  =\frac{1}{4}\int(\det e)\,\epsilon^{\mu\nu\kappa\lambda}\epsilon
_{\alpha\beta\gamma\delta}R_{\mu\nu}^{\ \ \alpha\beta}R_{\kappa\lambda
}^{\ \ \gamma\delta}\,d^{4}x\nonumber\\
&  =-\int\mathrm{d}^{4}x\,\frac{1}{4\sqrt{-g}}\,g_{\mu\rho}g_{\kappa\sigma
}\,\epsilon^{\mu\nu\kappa\lambda}\epsilon^{\alpha\beta\gamma\delta
}\,\,R_{\ \nu\alpha\beta}^{\rho}\,R_{\ \lambda\gamma\delta}^{\sigma},
\end{align}
where the minus sign follows from $\epsilon^{0123}=1$ and $\epsilon_{0123}%
=-1$. We have used the definition%
\begin{align}
R^{ab} &  \equiv\frac{1}{2}R_{\mu\nu}^{\quad ab}dx^{\mu}\wedge dx^{\nu
}\nonumber\\
&  =d\omega^{ab}+\omega^{ac}\wedge\omega_{c}^{\,\,\,b}%
\end{align}
where $\omega^{ab}=\omega_{\mu}^{\quad ab}dx^{\mu}$, which in turn is
determined by the zero torsion condition%
\begin{equation}
de^{a}=-\omega_{\quad b}^{a}\wedge e^{b},\qquad e^{a}=e_{\mu}^{a}dx^{\mu}%
\end{equation}
We note in passing, that this equation allows to compute $\omega_{\quad b}%
^{a}$ as function $e_{\mu}^{a}$ by first using the definition
\begin{equation}
de^{a}=-\frac{1}{2}C_{bc}^{\quad a}e^{b}\wedge e^{c}%
\end{equation}
and comparing using the symmetries to get%
\begin{equation}
\omega_{ab}=e^{c}\omega_{cab}=\frac{1}{2}e^{c}\left(  C_{abc}+C_{acb}%
-C_{bca}\right)
\end{equation}
To vary the action, we first use
\begin{equation}
\frac{1}{\sqrt{-g}}\,\delta\!\left(  \frac{1}{\sqrt{-g}}\right)  =-\frac{1}%
{2}g^{\mu\nu}\delta g_{\mu\nu}\,\frac{1}{(-g)},
\end{equation}
so that the first contribution is
\begin{align}
&  -\int\mathrm{d}^{4}x\,\frac{1}{2\sqrt{-g}}\,\delta g_{\mu\rho}%
g_{\kappa\sigma}\,\epsilon^{\mu\nu\kappa\lambda}\epsilon^{\alpha\beta
\gamma\delta}f(\phi)g\left(  \square\phi\right)  \,R_{\ \nu\alpha\beta}^{\rho
}\,R_{\ \lambda\gamma\delta}^{\sigma}\nonumber\\
&  =-2\int\mathrm{d}^{4}x\,\sqrt{-g}\,\delta g_{\mu\nu}\,f(\phi)g\left(
\square\phi\right)  \left(  R^{\nu\mu}R-2R^{\nu\rho\mu\beta}R_{\rho\beta
}+R^{\nu\rho\kappa\lambda}R_{\ \rho\kappa\lambda}^{\mu}-2R^{\nu\beta
}R_{\ \beta}^{\mu}\right)  ,
\end{align}
but terms proportional to $f(\phi)g\left(  \square\phi\right)  $ vanish by the
identity
\begin{equation}
0=R_{\eta\tau\alpha\beta}R^{\kappa\lambda\gamma\delta}\,\delta_{\nu
\kappa\lambda\gamma\delta}^{\mu\eta\tau\alpha\beta}.
\end{equation}
Next we consider variation of the curvature-squared terms:
\begin{equation}
-\int\mathrm{d}^{4}x\,\frac{1}{2\sqrt{-g}}\,g_{\mu\rho}g_{\kappa\sigma
}\,\epsilon^{\mu\nu\kappa\lambda}\epsilon^{\alpha\beta\gamma\delta}%
f(\phi)g\left(  \square\phi\right)  \,\delta R_{\ \nu\alpha\beta}^{\rho
}\,R_{\ \lambda\gamma\delta}^{\sigma}.
\end{equation}
Using
\begin{equation}
\delta R_{\ \nu\alpha\beta}^{\rho}=\nabla_{\alpha}\delta\Gamma_{\nu\beta
}^{\rho}-\nabla_{\beta}\delta\Gamma_{\nu\alpha}^{\rho},
\end{equation}
and integrating by parts:
\begin{align*}
& \int\mathrm{d}^{4}x\,\frac{1}{\sqrt{-g}}\,g_{\mu\rho}g_{\kappa\sigma
}\,\epsilon^{\mu\nu\kappa\lambda}\epsilon^{\alpha\beta\gamma\delta}%
\delta\Gamma_{\nu\beta}^{\rho}\,\nabla_{\alpha}\!\left(  f(\phi)g\left(
\square\phi\right)  R_{\ \lambda\gamma\delta}^{\sigma}\right) \\
& =\int\mathrm{d}^{4}x\,\frac{1}{\sqrt{-g}}\,g_{\mu\rho}g_{\kappa\sigma
}\,\epsilon^{\mu\nu\kappa\lambda}\epsilon^{\alpha\beta\gamma\delta}%
\delta\Gamma_{\nu\beta}^{\rho}\,R_{\ \lambda\gamma\delta}^{\sigma}%
\,\nabla_{\alpha}\left(  f(\phi)g\left(  \square\phi\right)  \right)  ,
\end{align*}
using the Bianchi identity. Writing
\begin{equation}
g_{\mu\rho}\delta\Gamma_{\nu\beta}^{\rho}=\frac{1}{2}\left(  \nabla_{\nu
}\delta g_{\mu\beta}+\nabla_{\beta}\delta g_{\mu\nu}-\nabla_{\mu}\delta
g_{\nu\beta}\right)  ,
\end{equation}
and integrating by parts (dropping antisymmetric contributions) gives
\begin{align}
& \int\mathrm{d}^{4}x\,\frac{1}{\sqrt{-g}}\,\epsilon^{\mu\nu\kappa\lambda
}\epsilon^{\alpha\beta\gamma\delta}R_{\kappa\lambda\gamma\delta}\,\nabla_{\nu
}\nabla_{\alpha}\left(  f(\phi)g\left(  \square\phi\right)  \right)  \,\delta
g_{\mu\beta}\nonumber\\
& =-\int\mathrm{d}^{4}x\,\frac{1}{\sqrt{-g}}\,\epsilon^{\mu\beta\kappa\lambda
}\epsilon^{\nu\alpha\gamma\delta}R_{\kappa\lambda\gamma\delta}\,\nabla_{\beta
}\nabla_{\alpha}\left(  f(\phi)g\left(  \square\phi\right)  \right)  \,\delta
g_{\mu\nu}.
\end{align}
The variation of the Einstein term and mimetic constraint is given by
\begin{equation}
\frac{1}{2}\int\mathrm{d}^{4}x\,\sqrt{-g}\,G^{\mu\nu}\delta g_{\mu\nu}%
-\int\mathrm{d}^{4}x\,\sqrt{-g}\,\lambda\,\partial_{\alpha}\phi\,\partial
_{\beta}\phi\,g^{\mu\alpha}g^{\nu\beta}\,\delta g_{\mu\nu}.
\end{equation}
Thus the equations of motion varying with respect to the metric are
\begin{align}
G_{\mu\nu} &  =2\lambda\,\partial_{\mu}\phi\,\partial_{\nu}\phi+\frac{2}%
{-g}\,R_{\kappa\lambda\gamma\delta}\,\nabla_{\beta}\nabla_{\alpha
}\!\big(f(\phi)g(\square\phi)\big)\,\epsilon^{\sigma\beta\kappa\lambda
}\epsilon^{\rho\alpha\gamma\delta}\,g_{\mu\rho}g_{\nu\sigma}\nonumber\\
&  \quad-\partial_{\nu}\phi\,\nabla_{\mu}\!\Big(f(\phi)g^{\prime}(\square
\phi)\,GB\Big)-\partial_{\mu}\phi\,\nabla_{\nu}\!\Big(f(\phi)g^{\prime
}(\square\phi)\,GB\Big)\nonumber\\
&  \quad+g_{\mu\nu}g^{\alpha\beta}\nabla_{\alpha}\!\Big(\partial_{\beta}%
\phi\,f(\phi)g^{\prime}(\square\phi)\,GB\Big),
\end{align}
where $GB$ denotes the Gauss--Bonnet combination%
\begin{equation}
GB=R_{\mu\nu}^{\quad\alpha\beta}R_{\alpha\beta}^{\quad\mu\nu}-4R_{\mu\nu
}R^{\mu\nu}+R^{2}%
\end{equation}
Finally, varying with respect to $\phi$ gives%
\begin{align}
\frac{1}{\sqrt{-g}}\partial_{\mu}\left(  2\sqrt{-g}g^{\mu\nu}\partial_{\nu
}\phi\lambda\right)   &  =f^{\prime}\left(  \phi\right)  g\left(  \square
\phi\right)  \left(  GB\right) \nonumber\\
&  +\frac{1}{\sqrt{-g}}\partial_{\mu}\left(  \sqrt{-g}g^{\mu\nu}\partial_{\nu
}\left(  f(\phi)g^{\prime}(\square\phi)GB\right)  \right)
\end{align}

\end{document}